\documentclass[aps,pra,twocolumn, 10pt, showpacs,superscriptaddress,groupedaddress,amsmath,amssymb]{revtex4-1} 

\usepackage{graphicx}
\usepackage{bm}
\usepackage{braket}
\usepackage{times}
\usepackage{subfig} 
\usepackage{caption}

\begin{document}


\title{Observation of Spin-Spin Fermion-Mediated Interactions between Ultracold Bosons}

\author{Hagai Edri}
\affiliation{Department of Physics of Complex Systems, Weizmann Institute of Science, Rehovot 76100, Israel }
\author{Boaz Raz}
\affiliation{Department of Physics of Complex Systems, Weizmann Institute of Science, Rehovot 76100, Israel }
\author{Noam Matzliah}
\affiliation{Department of Physics of Complex Systems, Weizmann Institute of Science, Rehovot 76100, Israel }
\author{Nir Davidson}
\affiliation{Department of Physics of Complex Systems, Weizmann Institute of Science, Rehovot 76100, Israel }
\author{Roee Ozeri}
\affiliation{Department of Physics of Complex Systems, Weizmann Institute of Science, Rehovot 76100, Israel }









\begin{abstract}

  Interactions in an ultracold boson-fermion mixture are often manifested by elastic collisions. In a mixture of a condensed Bose gas (BEC) and spin polarized degenerate Fermi gas (DFG), fermions can mediate spin-spin interactions between bosons, leading to an effective long-range magnetic interaction analogous to Ruderman-–Kittel-–Kasuya-–Yosida (RKKY) \cite{ruderman1954indirect,kasuya1956theory,yosida1957magnetic} interaction in solids. We used Ramsey spectroscopy of the hyperfine clock transition in a $^{87}$Rb BEC to measure the interaction mediated by a $^{40}$K DFG. By controlling the bosons density we isolated the effect of mediated interactions from mean-field frequency shifts due to direct collision with fermions. We measured an increase of boson spin-spin interaction by a factor of $\eta=1.45\pm 0.05^{stat} \pm 0.13^{sys}$ in the presence of the DFG, providing a clear evidence of spin-spin fermion mediated interaction. Decoherence in our system was dominated by inhomogeneous boson density shift, which increased significantly in the presence of the DFG, again indicating mediated interactions. We also measured a frequency shift due to boson-fermion interactions in accordance with a scattering length difference of $a_{bf_2}-a_{bf_1}=-5.36 \pm 0.44^{stat} \pm 1.43^{sys}~a_0$ between the clock-transition states, a first measurement beyond the low-energy elastic approximation \cite{cote1998potassium,dalgarno1965spin} in this mixture. This interaction can be tuned with a future use of a boson-fermion Feshbach resonance. Fermion-mediated interactions can potentially give rise to interesting new magnetic phases and extend the Bose-Hubbard model when the atoms are placed in an optical lattice.
  \end{abstract}

\maketitle

Mediated interactions are at the heart of modern physics. In the standard model, every interaction has a bosonic mediator. In solid state physics, electrons can form cooper pairs by interacting via phonons, resulting in BCS superconductivity \cite{bardeen1957theory}, or serve as mediators for interactions between magnetic impurities, leading to  Ruderman–-Kittel-–Kasuya–-Yosida (RKKY) interaction \cite{ruderman1954indirect,kasuya1956theory,yosida1957magnetic}. This interaction is a spin-spin long range mediated interaction, it was observed between spin impurities in metals \cite{bloembergen1953nuclear,pruser2014interplay,kittel1969indirect} and quantum dots in semiconductors \cite{simon2005ruderman,craig2004tunable,PhysRevLett.96.017205}. The mechanical effects of a mean-field spinless analog of RKKY were recently observed in ultracold atoms \cite{desalvo2019observation}. Theoretical studies proposed it should exist in graphene \cite{saremi2007rkky,sherafati2011rkky,allerdt2017competition},  and in an ultracold Bose-Fermi mixture \cite{de2014fermion,santamore2008fermion}. 

Spin-spin magnetic interactions are ubiquitous in physics, NMR, Kondo effect and Ising model are prime examples. Magnetic interactions lead to different magnetic phases in metals, such as ferromagnetic and anti-ferromagnetic ordered phases. Interactions in ultracold atomic gases can be also regarded as spin-spin interactions in the mean field approximation.  Effective magnetic interactions were observed in spinor BEC \cite{sadler2006spontaneous,marti2014coherent} and fermions in optical lattices \cite{mazurenko2017cold,salomon2019direct}. In a quantum degenerate Bose-Fermi mixture, fermion mediated interactions have been predicted to create spin-spin long range interaction between bosons that can be tuned with a Feshbach resonance \cite{de2014fermion,santamore2008fermion}. Mediated magnetic interactions may lead to interesting new magnetic phases of matter \cite{nishida2010phases,suchet2017long,caracanhas2017fermi} and can extend the Bose-Hubbard model when the atoms are placed in an optical lattice \cite{de2014fermion}. 

Unlike super-exchange interactions, which require the atoms to be at close contact, fermion mediated interactions have a range which exceeds the Fermi wavelength of the mediating DFG. Other forms of long range interactions in ultracold gases have been observed in gases of highly magnetic atoms \cite{baier2016extended,kadau2016observing,stuhler2005observation}, Rydberg atoms \cite{bendkowsky2009observation,urban2009observation,dudin2012strongly}, and in polar molecules \cite{yan2013observation,park2015ultracold}. Mediated interactions were previously seen in Bose-Einstein condensates coupled to an optical cavity \cite{mottl2012roton,norcia2018cavity,PhysRevLett.122.193601,vaidya2018tunable}, where photons act as mediators.

Fermion mediated interactions can be intuitively understood. A boson immersed in a Fermi sea induces a perturbation to the fermion density that oscillates with a typical length inversely proportional to the Fermi wave-number $k_f$, known as Friedel oscillations \cite{friedel1952xiv}. When another boson interacts with the deformed Fermi sea, interaction is mediated between the bosons. Spin-dependent coupling of the bosons to the Fermi sea leads to spin-spin magnetic mediated interactions.

In a quantum degenerate mixture of Bose-Einstein condensate (BEC) and a polarized degenerate Fermi gas (DFG), atoms interact via $s$-wave collisions, characterized by scattering lengths $a_{bb}$ and $a_{bf}$ for boson-boson and boson-fermion interactions respectively. Boson-boson $s$-wave collisions in a BEC cause a mean field energy shift $E_{bb}=2\pi\frac{\hbar^2}{m_b}a_{bb}n_b$ \cite{harber2002effect} where  $\hbar$ is the reduced Planck constant, $n_b, m_b$ are boson density and boson mass respectively. For light fermions ($m_f\ll m_b$ ,where $m_f$ is the fermion mass), an exchange of fermionic excitation between bosons results in a long range potential, analogous to RKKY interaction in solids $V(R)\propto- a_{bf}^2\frac{\sin\left(2k_fR\right)-2k_fR\cos\left(2k_f R\right)}{\left(2k_fR\right)^4}$ \cite{de2014fermion,santamore2008fermion} where $R$ is distance between bosons. In a mean field approximation, the energy shift due to this potential is $E_{med} = \frac{2\pi\hbar^2}{m_b}a_{med}n_b$ where $a_{med}=\frac{\xi}{2\pi}\frac{\left(m_b+m_f\right)^2}{m_b m_f}\left(6\pi^2n_f\right)^{1/3}a_{bf}^2$ is the mediated interaction scattering length, $n_f$ is the fermion density, and $\xi$ is a dimensionless parameter that characterizes the strength of fermion-mediated interaction, predicted to be $\pi^3$ \cite{de2014fermion} or $1$ \cite{santamore2008fermion}. We note that this potential is expected to have corrections beyond the mean-field approximation since it decays as $1/R^3$ and partial-waves of higher order contribute even at zero temperature. We also note that the bosons are in a BEC at the ground state of the system, and the healing length of the BEC is longer than the typical length scale of the potential $1/2k_f$.

\begin{figure*}
        \centering
		\begin{minipage}[b]{1\textwidth}
			\centering
			\subfloat[]{\includegraphics[width=0.5\textwidth]{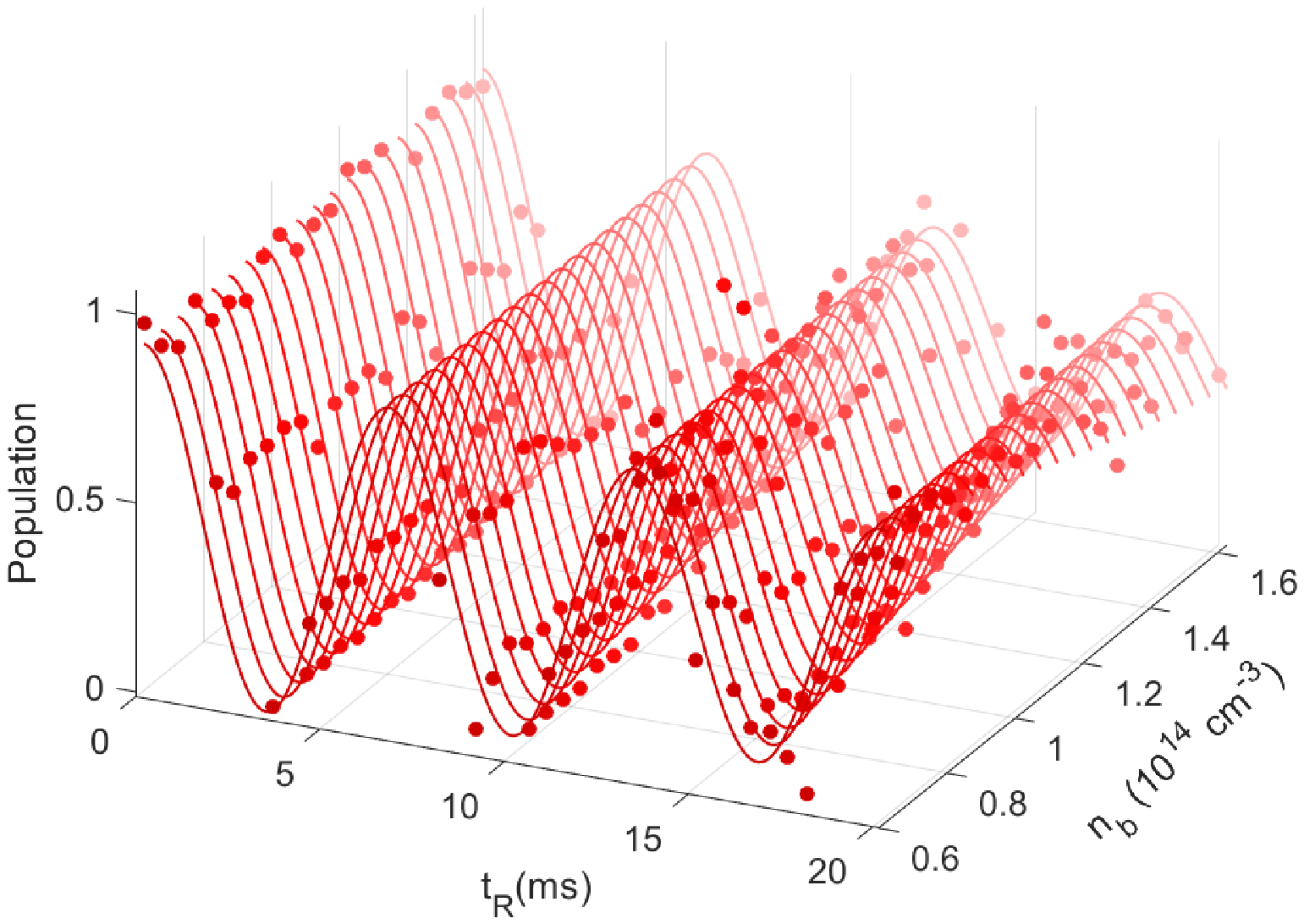} }			\subfloat[]{\includegraphics[width=0.5\textwidth]{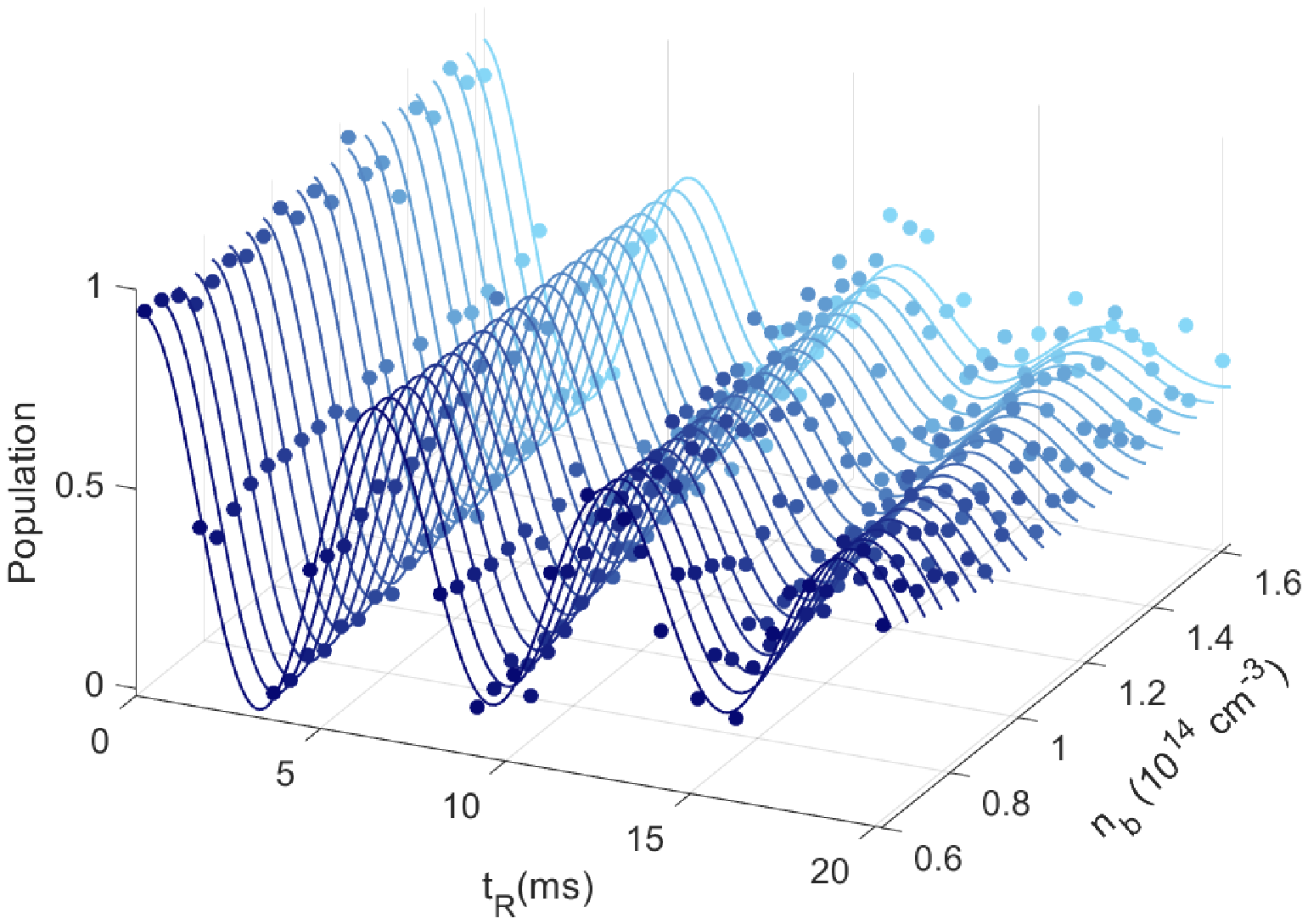} }
		\end{minipage}		
		\caption{Ramsey spectroscopy -  $^{87}$Rb $F=2$ population vs. Ramsey time measured at different boson densities for (a) bosons only and (b) a mixture of bosons and fermions. Data points are averaged for different boson densities to avoid cluttering (one standard deviation errors are comparable to marker size). The solid lines are the result of a single 2D fit to all data points (see Eq. (2) in the text). We measured a boson density shift of $\left(\frac{\mathrm{d}\Delta{f}}{\mathrm{d}n_b}\right)_b = 10.77\pm 0.27^{stat} \pm 1.69^{sys} \times10^{-14}$~Hz$\cdot$cm$^3$ for bosons (a) and $\left(\frac{\mathrm{d}\Delta{f}}{\mathrm{d}n_b}\right)_{b+f} = 15.53 \pm 0.39^{stat} \pm 2.37^{sys} \times10^{-14}$~Hz$\cdot$cm$^3$  for the boson-fermion mixture (b), this change is a clear effect of spin-spin mediated interaction. The signal decays when increasing boson density in both cases, and to a larger extent when fermions are present, another indication of mediated interactions.}
		\label{fig:data}
\end{figure*}

In our experiment, a BEC of $\sim 5\times10^5$  $^{87}$Rb atoms was in thermal contact with a DFG of $\sim 1.4\times10^5$ $^{40}$K atoms in a crossed dipole trap of frequencies $\omega_{x,y,z}=2\pi\times \left(\ 27, 39, 111\right)$ Hz ($\omega_{x,y,z}=2\pi\times\left(\  29, 49, 182 \right)$ Hz) for bosons (fermions) at a temperature of $\sim90$ nK ($T/T_c\sim0.55$ and $T/T_f\sim0.35$), with mean BEC (DFG) Thomas-Fermi (Fermi) radius of $\sim11$ $\mu$m ($\sim26$ $\mu$m). The fermions are spin polarized in state $\ket{F=9/2,m_f=-9/2}$. 
When calculating fermion mediated interaction, we assume that the bosons are stationary ($m_f\ll m_b$) and the change in their energy after a boson-fermion collision is negligible, while in our system the mass ratio is $\sim 2$. However, the  fermions involved in the process are close to the Fermi momentum and the BEC has essentially zero momentum. After an exchange of momentum $q\ll k_f$, the change in energy for fermions $\Delta \epsilon_f$ is much larger than for bosons $\Delta \epsilon_b$, since $\frac{\Delta \epsilon_f}{\Delta \epsilon_b}=\frac{k_f}{q}\frac{m_b}{m_f}\gg1$.

We used microwave spectroscopy to measure the $\ket{F=1,m_f=0}=\ket{1} \leftrightarrow \ket{F=2,m_f=0}=\ket{2}$ hyperfine clock transition frequency in $^{87}$Rb with and without $^{40}$K atoms. The clock transition frequency shift includes three contributions - boson-boson interaction $\Delta{f_{bb}}$, boson-fermion interaction $\Delta{f_{bf}}$ and mediated interaction $\Delta{f_{med}}$. 
\begin{equation}
\begin{split}
\Delta{f} =\underbrace{2 \frac{\hbar}{m_b}(a_{bb_2}-a_{bb_1})n_b}_{\Delta{f_{bb}}}&+\underbrace{\hbar(\frac{1}{m_b}+\frac{1}{m_f})(a_{bf_2}-a_{bf_1})n_f}_{\Delta{f_{bf}}}\\
+&\underbrace{2\frac{\hbar}{m_b}(a_{med_2}-a_{med_1})n_b}_{\Delta{f_{med}}}
\end{split}
\end{equation}
$\Delta{f_{bb}}$ and $\Delta{f_{med}}$ are proportional to boson density, while $\Delta{f_{bf}}$ is independent of it. To distinguish between the three contributions, we measured frequency shifts for different boson densities, with and without fermions. With only bosons present, we measured $\Delta{f_{bb}}$, the change in frequency per boson density $\left(\frac{\mathrm{d}\Delta{f}}{\mathrm{d}n_b}\right)_b$ which is a direct measure of boson-boson interaction. When we added fermions, we measured a constant term $\Delta{f_{bf}}$ that does not depend on boson density and $\Delta{f_{med}}$, which causes change in frequency that is proportional to boson density. A change in the slope $\left(\frac{\mathrm{d}\Delta{f}}{\mathrm{d}n_b}\right)_{b+f}$ (measurement with bosons and fermions) compared to $\left(\frac{\mathrm{d}\Delta{f}}{\mathrm{d}n_b}\right)_b$ (measurement with bosons only) is a clear indication of fermion mediated interactions between bosons in our mixture.

The clock transition states are first-order magnetic insensitive at zero magnetic field and thus also have the same electronic triplet and singlet components when colliding with another atom. This renders the difference in scattering lengths $a_{bf_2}-a_{bf_1}$ between these two states small as compared with other transitions \cite{cote1998potassium,dalgarno1965spin} and thus decreases the mediated interaction shift $\Delta{f_{med}}$ and the boson-fermion shift $\Delta{f_{bf}}$.

\begin{figure*}
    \label{frequency shift}
	\centering
	\begin{minipage}[b]{1\textwidth}
		\centering
		\subfloat[]{\includegraphics[width=0.5\textwidth]{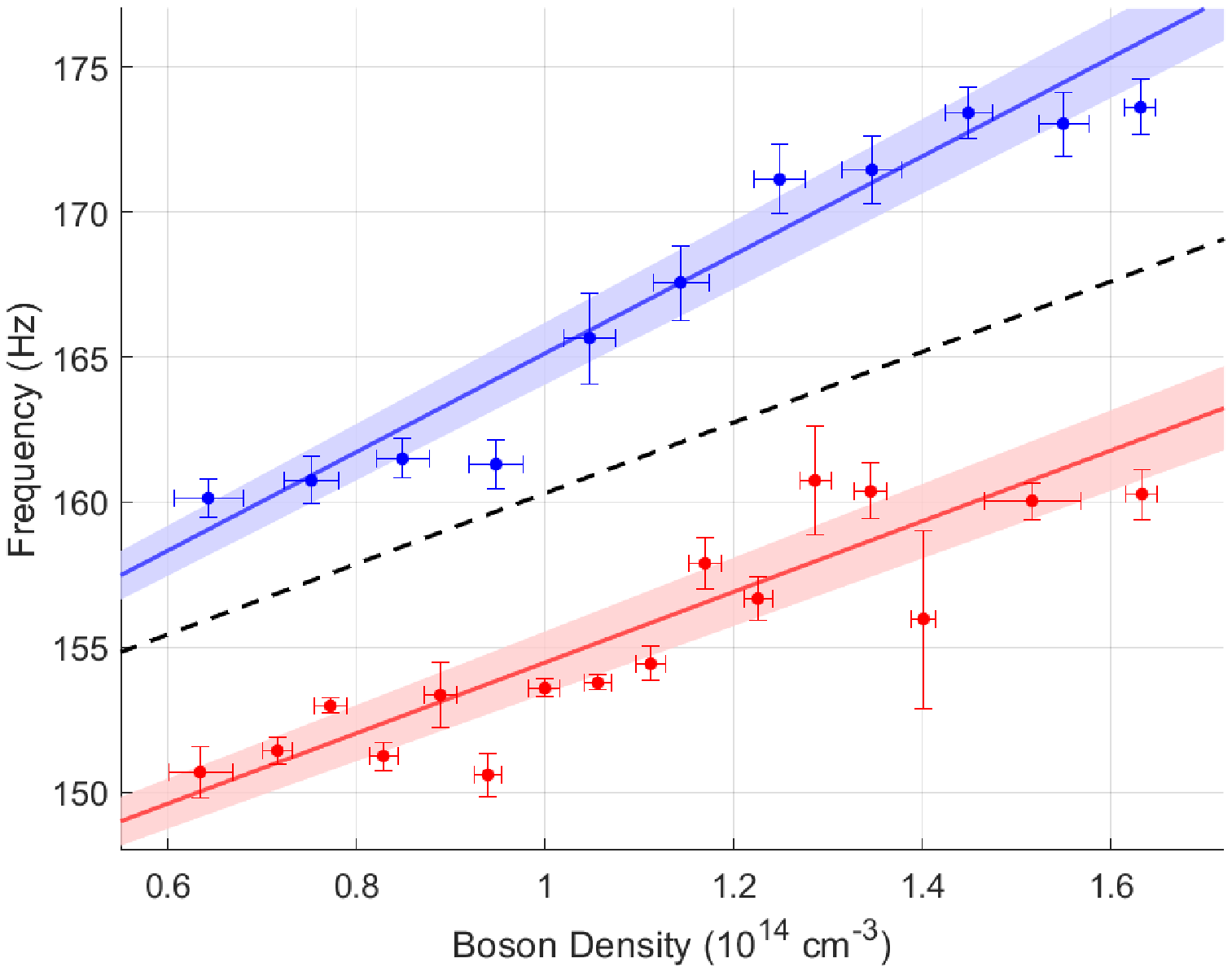}}
		\subfloat[]{\includegraphics[width=0.5\textwidth]{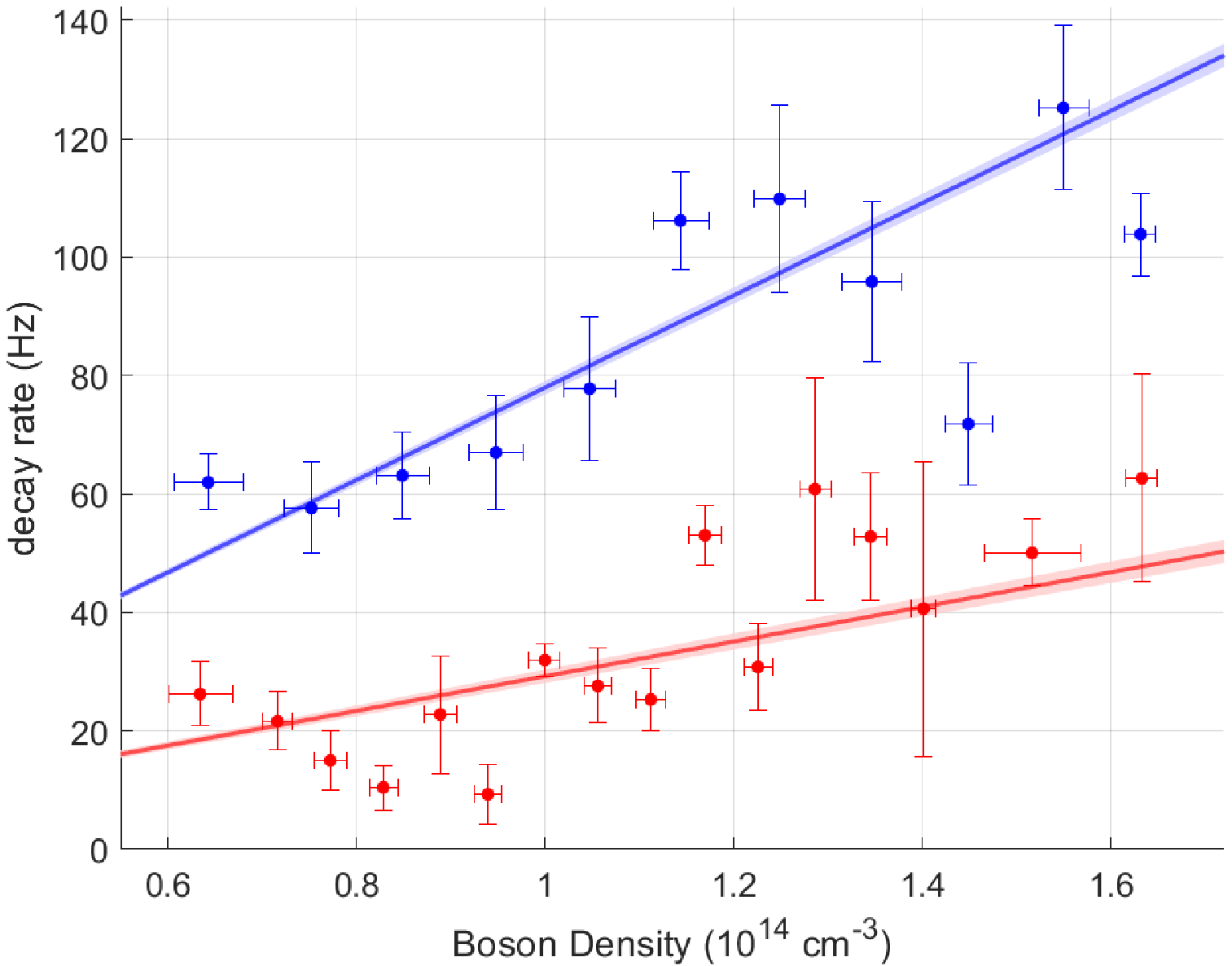}}
	\end{minipage}		
	\caption{Frequency shift (a) and decay rate (b)  for different boson densities for bosons only (red) and and a mixture of bosons and fermions  (blue). The lines are extracted from a 2D fit to all data points (shaded area denotes one standard deviation), the filled circles are extracted from fits to averaged data for different boson densities (density errors are statistical standard deviation from binning). A dashed black line is shown in (a) parallel to red line and  crosses the blue line  at zero density. (a) A constant frequency shift between the two lines (i.e. independent of boson density) is a measure of boson-fermion interaction with  $a_{bf_2}-a_{bf_1} = -5.36 \pm 0.44^{stat} \pm 1.43^{sys}~a_0$. A change in the slope by a factor $\eta=1.45\pm 0.05^{stat} \pm 0.13^{sys}$ is a clear effect of fermion mediated interaction. Both of these effects are insensitive to any systematic error that is common to measurements with/without fermions (i.e. light shift, magnetic field fluctuations and errors in estimating the boson density). (b) The decay rate increases with boson density in both cases (with/without fermions), while the increase is faster when fermions are present, indicating mediated interactions.}
	\label{fig:lines}
\end{figure*}

Following a Ramsey experiment of variable duration $t_R$, we released the atoms from the dipole trap. After a long time-of-flight (20~ms for the bosons and 12~ms for the fermions) we used absorption imaging to measure $P=N_2/N_{tot}$, the boson population in state $\ket{2}$, and the BEC chemical potential and the fermion fugacity, which we used to determine boson and fermion densities in each run. The BEC chemical potential $\mu_b$ is determined from a bimodal fit function to the column density  \footnote{For further discussion see supplementary material} of the BEC in the tightly confined direction of our trap ($\omega_z > \omega_{x,y}$) after expansion, from that we calculate the average boson density $\bar{n}=\frac{4}{7}n_0 =\frac{\mu_b}{7\pi\hbar^2 a_{bb}}$, here $n_0$ is the peak density \cite{RevModPhys.71.463}. 

The presence of the fermions may distort our boson density measurements as the chemical potential of the BEC changes due to boson-fermion interaction \cite{santamore2008fermion}. This change is $\sim \frac{1}{7}\mu_b$, it is uniform across the BEC (since the size of the fermions cloud is larger due to Fermi pressure), and  fairly constant throughout our measurement. During the release from the trap the interaction energy is divided between bosons and fermions as the clouds expand. To verify that it does not introduce a significant systematic error to our density evaluation, we compared the chemical potential of BEC’s with similar atom number (determined from absorption imaging) with and without fermions. We did not observe a change in measured chemical potential after expansion in the presence of fermions \cite{Note1}.

We controlled the boson density during state preparation, by transferring a predetermined fraction of $^{87}$Rb atoms to state $\ket{2}$, and removing this fraction from the trap using a resonant laser pulse. Measurements without fermions were performed in an identical fashion to obtain similar boson densities, expelling the fermions at the end of preparation stage, immediately before the Ramsey measurement.

Our results are shown in Figure~\ref{fig:data}. Red (Blue) filled circles show the measured populations without (with) fermions. Solid lines are the result of a maximum likelihood fit to the data using the fit function: 
\begin{equation}
P(n_b,t_R)=A e^{-g_c n_b t_R} \cos\left((\delta+g n_b) t_R \right)+C.
\end{equation}
Here, $A$ is the contrast, $g_c$ is decoherence rate per boson density, $\delta$ is detuning between the MW driving field and the transition, $g=\frac{\mathrm{d}\Delta{f}}{\mathrm{d}n_b}$ is the boson density frequency shift, and $C$ is a constant term to account for a finite population in state $\ket{2}$. We extracted three frequency shifts from the fits. The boson density shift in the absence of fermions, $\Delta{f_{bb}}$, the mean-field shift due to direct boson-fermions collisions, $\Delta{f_{bf}}$, and the boson density shift which was mediated by the presence of fermions, $\Delta{f_{med}}$.

The measured frequencies and decay rates are shown in Figure~\ref{fig:lines}. Red (Blue) lines show the result from our fit function (eq. (2)) without (with) fermions. Filled circles show frequencies taken from a fit function $P(t_R) = A e^{-\Gamma t_R} \cos\left(2\pi f t_R \right)+C$ to averaged data for different densities. Here, $A$ is the contrast, $\Gamma$ is decay rate, $f$ is oscillation frequency, and $C$ is a constant term to account for a finite population in state $\ket{2}$. Since the light shifts, local-oscillator detuning and Zeeman shifts are independent of the presence of fermions, the difference in frequency between measurements with and without fermions is a result of boson-fermion density shift $\delta_{b+f}-\delta_{b}=\Delta{f_{bf}}$. Our frequency measurements  (Figure~\ref{fig:lines} (a)) show a constant shift with respect to boson density due to boson-fermion interaction $\frac{\mathrm{d} \Delta f_{bf}}{\mathrm{d}n_f}= 6.57 \pm 0.53^{stat} \pm 1.75^{sys}\times10^{-13}$~Hz$\cdot$cm$^3$ from which we calculate a boson-fermion scattering length difference $a_{bf_2}-a_{bf_1}=-5.36 \pm 0.44^{stat} \pm 1.43^{sys} ~a_0$ where $a_0$ is Bohr radius (all errors reported are of one standard deviation \cite{Note1}). To the best of our knowledge, the scattering length for $^{87}$Rb$-^{40}$K collisions is only previously known from Feshbach spectroscopy measurement \cite{ferlaino2006feshbach}. Our measured value is a correction to the low-energy elastic approximation \cite{cote1998potassium,dalgarno1965spin} and can be used to calibrate inter atomic potential calculations.

When measuring without fermions, $g$ is a direct measurement of boson-boson interaction. We measure a shift  $\left(\frac{\mathrm{d}\Delta{f}}{\mathrm{d}n_b}\right)_b= 10.77\pm 0.27^{stat} \pm 1.69^{sys} \times10^{-14}$~Hz$\cdot$cm$^3$ that corresponds with a difference in scattering lengths $a_{bb_2}-a_{bb_1}=-1.40 \pm 0.04^{stat} \pm 0.22^{sys}~a_0$,  in agreement with previously measured values \cite{fertig2000measurement,PhysRevLett.85.3117} (after taking into account a factor of 2 for BEC statistics \cite{harber2002effect} and uncertainties in atom numbers). 

When adding fermions we measure a change in the slope $\left(\frac{\mathrm{d}\Delta{f}}{\mathrm{d}n_b}\right)_{b+f}$ by a factor $\eta=1.45\pm 0.05^{stat} \pm 0.13^{sys}$ compared to a our measurement without fermions, which is a clear indication of spin-spin mediated interactions. The mediated interaction frequency shift is $\frac{\mathrm{d}\Delta f_{med}}{\mathrm{d}n_b} = 4.76\pm 0.47^{stat} \pm 0.75^{sys} \times10^{-14}$~Hz$\cdot$cm$^3$. Our measurements were done in a randomized fashion, interlacing different boson densities  with/without fermions. Any systematic errors we have regarding our estimation of boson densities is common in both measurements. The slope ratio $\eta=\left(\frac{\mathrm{d}\Delta{f}}{\mathrm{d}n_b}\right)_{b+f}/\left(\frac{\mathrm{d}\Delta{f}}{\mathrm{d}n_b}\right)_b = 1+\frac{\Delta f_{med}}{\Delta f_{bb}}$ only depends weakly on the fermion density ($\eta -1 \propto n_f^{1/3}$). The fermion density in our measurements is $(8.6\pm 2) \times10^{12}$ cm$^{-3}$, we divided our data for different fermion densities and fitted to eq. (2), but did not see significant change in $\eta$. When using our measured values for $a_{bf_2}-a_{bf_1}$ and a previously known value for $a_{bf_1}=-185 \pm 7 ~a_0$ \cite{ferlaino2006feshbach} we calculate the expected mediated interaction shift and find $\xi=1.02 \pm 0.10^{stat} \pm 0.40^{sys}$, in good agreement with $\xi = 1$ theory \cite{santamore2008fermion} and a recent mechanical measurement \cite{desalvo2019observation}.

The measured decay rates are shown in Figure~\ref{fig:lines} (b).
Our coherence time is limited by inhomogeneous density shifts in the BEC, previous Ramsey measurements with thermal clouds and at lower densities have shown longer coherence times ($>0.5$ s). Our results show  (Figure ~\ref{fig:lines} (b)) a clear dependence of the decay rate on the boson density. When adding the fermions the decay is faster and it also increases more rapidly with boson density by a factor of $2.3\pm 0.5$. For a BEC immersed in a homogeneous cloud of fermions, the coherence decay should not be affected by direct collisions with the fermions. However, fermion mediated interactions should affect the decay as it changes the effective scattering length between the bosons in an in-homogeneous manner. We also note that, besides inhomogeneous density broadening, the spin-dependent coupling of bosonic superposition to a bath of fermions is expected to lead to increased decoherence, as an intrinsic aspect of mediated interactions. The change in the derivative of the decay rate with respect to boson density is another indication of mediated interactions in the mixture. 

In summary, we have performed a first spectroscopic measurement of spin-spin fermion mediated interaction in a quantum degenerate Bose-Fermi mixture. We measure the long wavelength limit of the response function of bosons to a density perturbation caused by the fermions \cite{santamore2008fermion,de2014fermion}, a mean-field effect of fermion-mediated interaction in our mixture. We used precision spectroscopy to measure frequency shifts of internal states of the bosons due to mediated interactions, the accuracy of our measurement enables us to measure the spin dependence of mediated interaction which is only $\sim5\%$ of the interaction itself. We measured an interaction of a $\sigma_z\cdot \sigma_z$ type, which is relevant in Ising model and spin phases, among others. Our method can be used to measure other types of mediated interactions in mixtures. It is a first step towards realizing long-range mediated interaction between atoms in a lattice, extending the known Bose-Hubbard model beyond nearest neighbor interactions. While the effect is small in our measurement, it can be increased significantly using an inter-species Feshbach resonance or by probing a different transition where the scattering length difference is larger.

\begin{acknowledgments}
This work was supported by the Israeli Science Foundation and the Israeli Ministry of Science Technology and Space.
\end{acknowledgments}

\bibliographystyle{apsrev4-1}
\bibliography{mediated_interaction}

\end{document}


\heading{Supplementary Material for Observation of Spin-Spin Fermion-Mediated Interactions between Ultracold Bosons}
\begin{center} Hagai Edri, Boaz Raz, Noam Matzliah, Nir Davidson and Roee Ozeri\end{center}
\begin{center} Department of Physics of Complex Systems, Weizmann Institute of Science, Rehovot 76100, Israel\end{center}


\subsection*{Experimental Setup}
\begin{figure}[ht]
   \centering
    \includegraphics[width=0.75\textwidth]{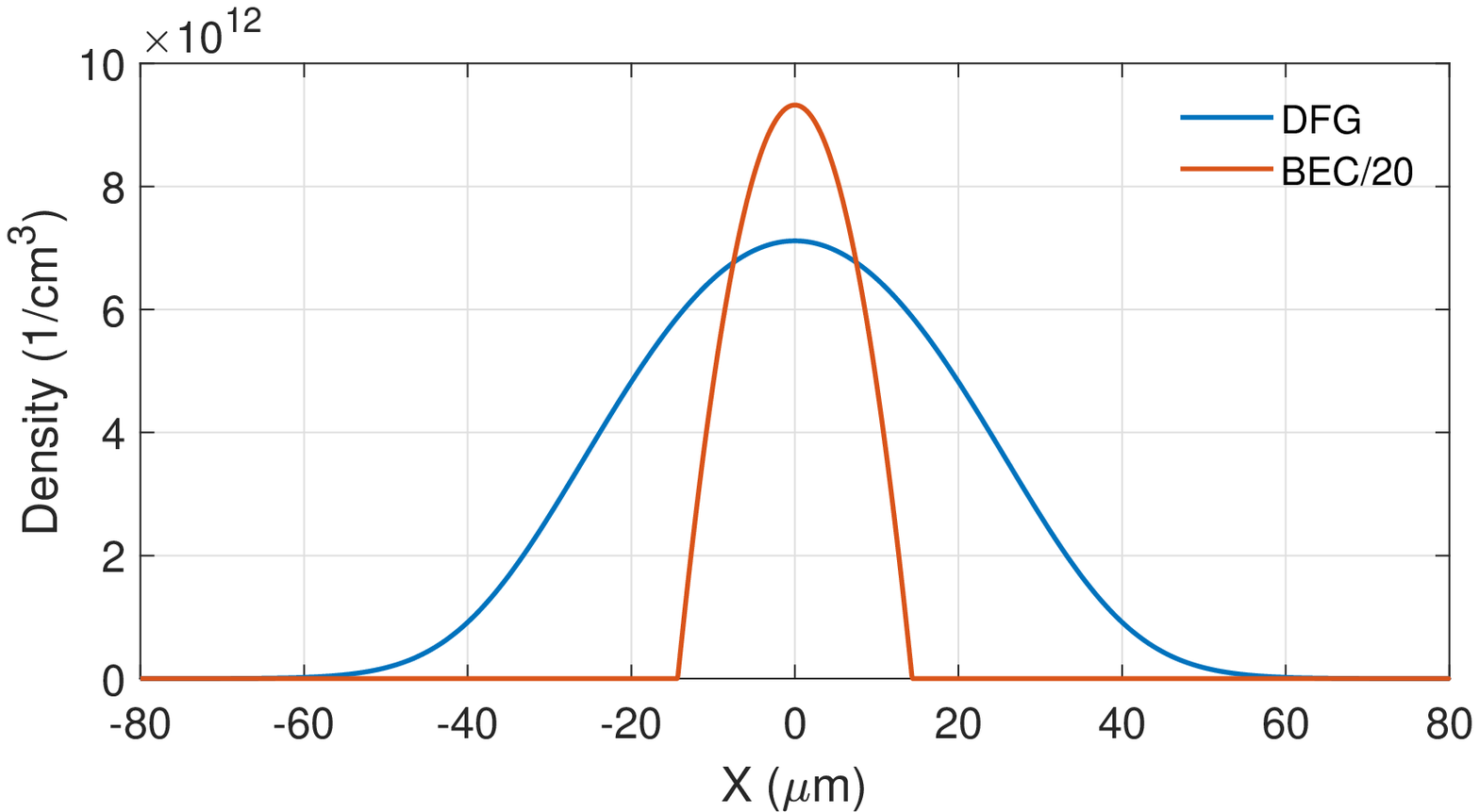}
    \caption{\label{fig1a} Calculated boson and Fermion density distributions in the trap – A cut through the trap center for boson (red) and fermion (blue) densities. The boson density is reduced by a factor of 20 for clarity. Where the boson density changes by 50\%, the fermion density changes only by $\sim10\%$ . }
\end{figure}

In our setup we simultaneously trap $^{87}$Rb and  $^{40}$K atoms  (bosons and fermions, respectively) in a crossed dipole trap (1064nm wavelength). At the start of our measurement we get a degenerate Fermi gas with $T/T_f\sim0.35$ and a BEC with condensate fraction $>70\%$ at a temperature of $\sim90$ nK. The trapping frequencies in the final trap are $\omega_{x,y,z}=2\pi\times \left(\ 27, 39, 111\right)$ Hz ($\omega_{x,y,z}=2\pi\times\left(\ 29, 49, 182 \right)$ Hz) for bosons (fermions). In the trap the mean BEC (DFG) Thomas-Fermi (Fermi) radius of $\sim11$ $\mu$m ($\sim26$ $\mu$m) A calculation of the boson and fermion densities in-situ is presented in Figure \ref{fig1a}.

\subsection*{Measuring boson density}
To measure density we release the atoms from the trap and image the atoms after 12~ms (fermions) and 20~ms (bosons) time of flight. After integrating along one axis we fit the bosons column density along the strong trapping direction with a bimodal distribution function $f(x)=Ae^{-\frac{x^2}{\sigma^2_x}}+n_c\cdot \max\left[0,\left(1-\frac{x^2}{R^2_x}\right)^2\right]$, here $\sigma_x$ ($R_x$) are parameters for the cloud size of the thermal (condensed) part, and $A$, $n_c$  are the amplitudes. We calculate the chemical potential $\mu_b=\frac{1}{2}m_b\frac{R_x^2}{t^2}$, here $m_b$ is boson mass, and $t$ is the time of flight. The peak density is $n_{0}=\mu\frac{m}{4\pi\hbar^2a_{bb}}$, and the average boson density is given by $n_b=\frac{4}{7}n_{0}$, our typical error in density from the fit function is $5\%$. The fermion image is fitted with a 2D Thomas-Fermi distribution from which we extract the fugacity and density.

\begin{figure}[tb]
   \centering
    \includegraphics[width=0.75\textwidth]{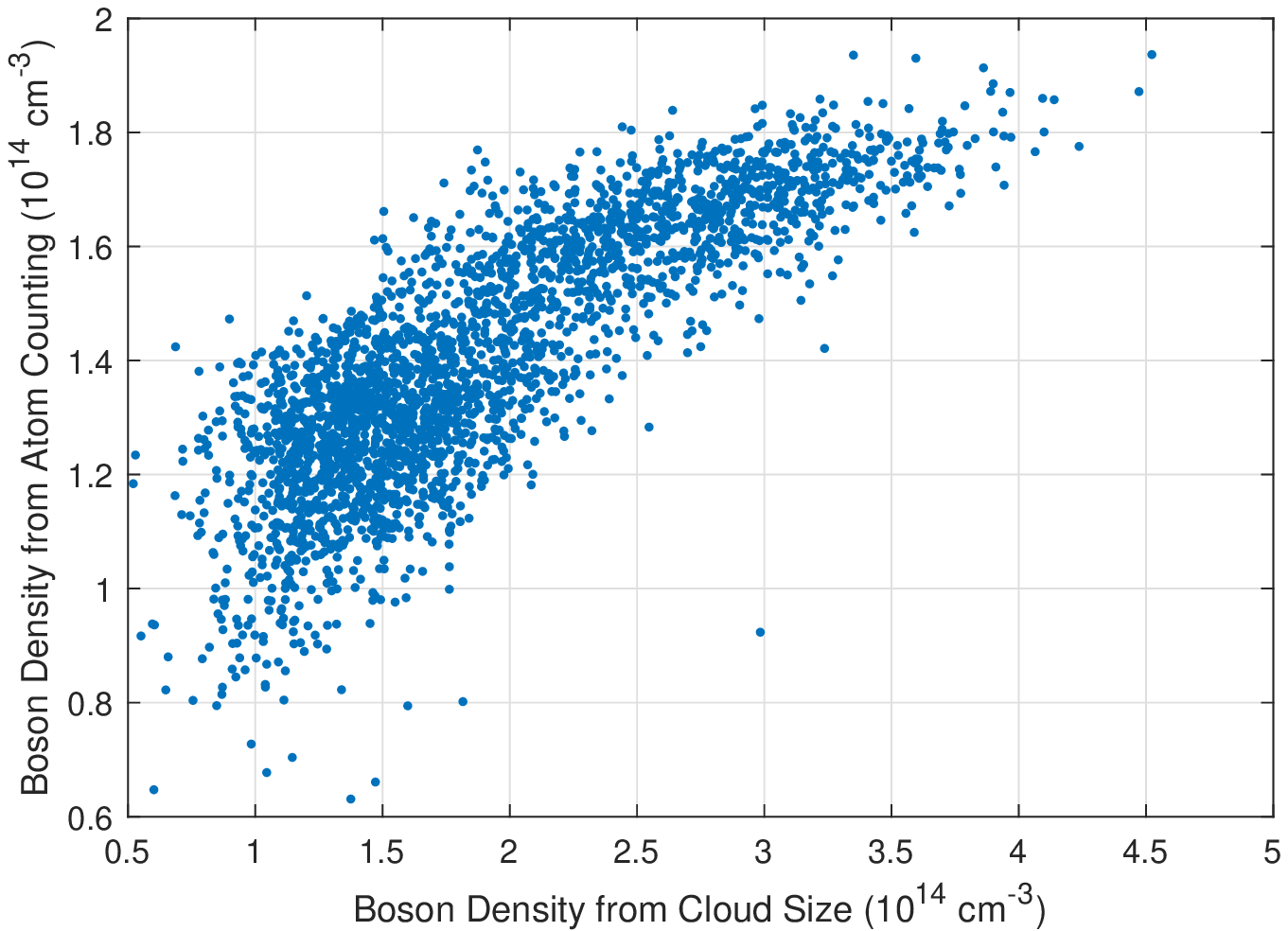}
    \caption{\label{fig1}  Comparison of boson density calculated from measuring cloud size after time of flight and calculated from counting the atoms in the absorption image.}
\end{figure}

We compare the boson density calculated from the size of the cloud to the density measured by counting the atoms from the optical density in the absorption image (Figure \ref{fig1}). We can see that the relation between the two is linear and agrees at the $50\%$ level at low boson densities and then the density calculated from the number of atoms saturates. This saturation can be explained by the high OD we have in our BEC at high boson densities.

Our measurement of boson density relies on measuring the size of the cloud after a long time of flight. To align our imaging system and check our resolution, we imaged the cloud in situ where its size is $\sim20$ times smaller than what we typically measure after time of flight. The pixel size in our camera is calibrated from measuring the free fall of the cloud after release and comparing it to gravitational acceleration. Our density measurement is affected by the sharpness of the edges of the BEC. In a similar way, our measurement of the fermion density is also susceptible to imaging errors at the edges where the signal is small due to low density. From in situ images of both clouds we estimate the overlap of both clouds to be $>90\%$.

\begin{figure}[tb]
 \centering
    \begin{subfigure}{.45\textwidth}
        \includegraphics[width=0.9\textwidth]{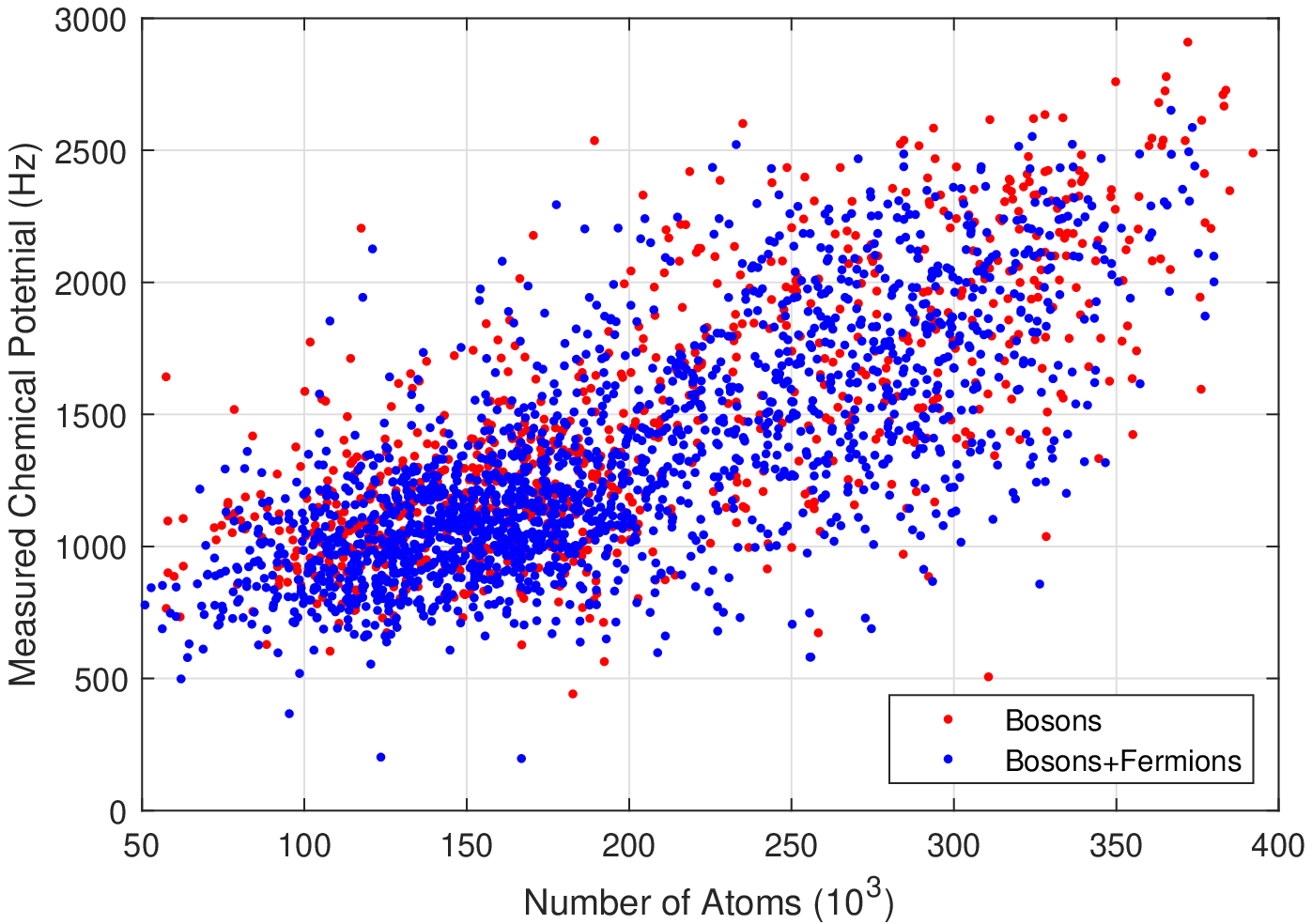}
        \caption{}
    \end{subfigure}
    \begin{subfigure}{.45\textwidth}
        \includegraphics[width=0.9\textwidth]{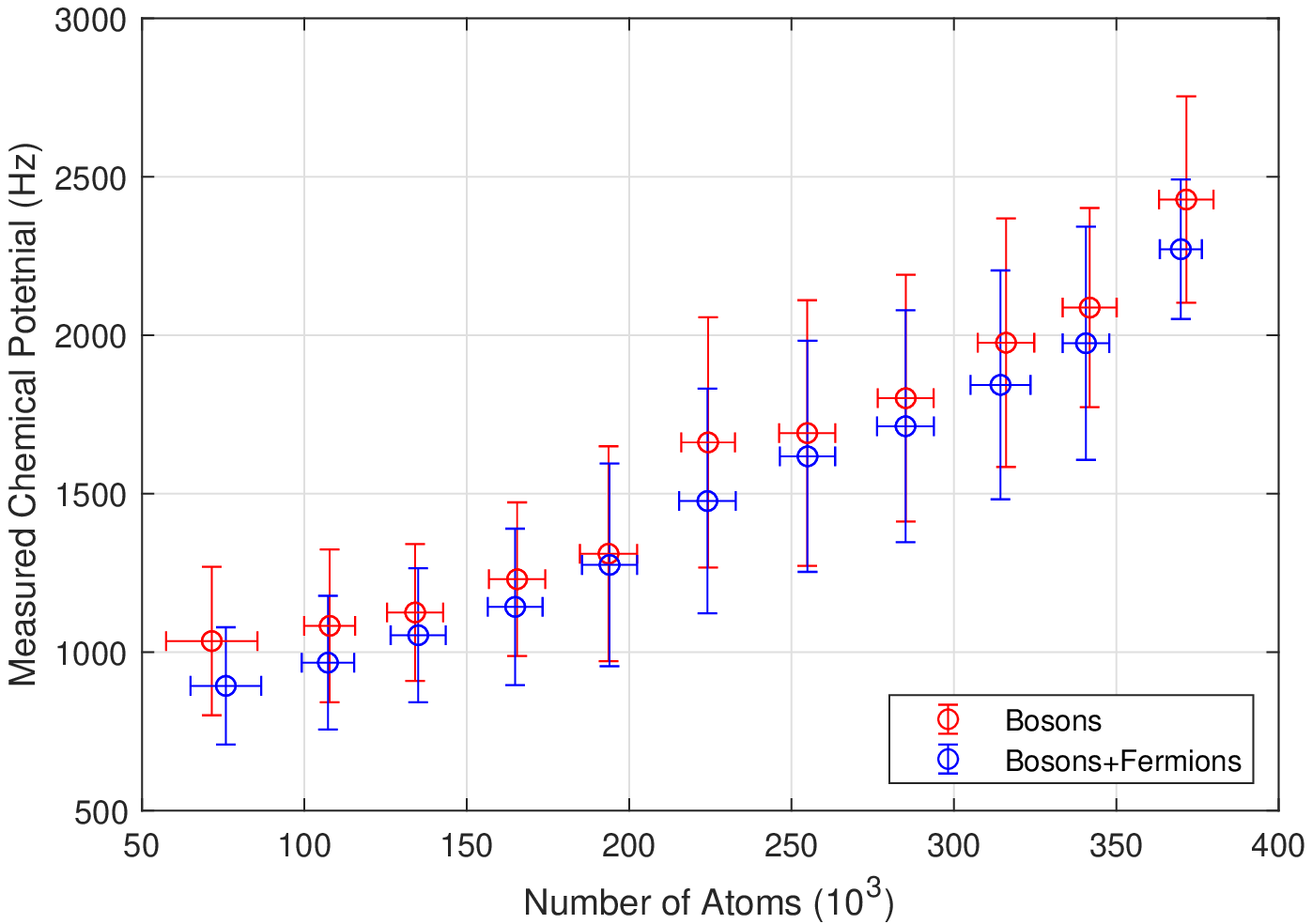}
        \caption{}
    \end{subfigure}
\caption{\label{fig2} Comparison of chemical potential with (blue) and without (red) fermions. Raw data (a) and averages for different number of atoms
(b), error bars show one standard deviation. There is no statistically significant difference in
measured chemical potential after expansion with/without fermions. Showing that our measured boson density is undisturbed by the presence of fermions.}
\end{figure}

The presence of the fermions changes the chemical potential of the BEC due to boson-fermion interaction. The BEC chemical potential is $\mu_b=\mu_{bb}+\mu_{bf}$, here $\mu_{bb}=\frac{4\pi\hbar^2}{m_b} n_0 a_{bb}$ ($\mu_{bf}=\frac{2\pi\hbar^2}{m_b} \left( 1+\frac{m_b}{m_f}\right) n_f a_{bf}$) is the energy due to boson-boson (boson-fermion) interaction, $m_b$ ($m_f$) is the boson  (fermion) mass, $n_b$ ($n_f$) is the boson (fermion) density, $a_{bb}$ ($a_{bf}$) is the boson-boson (boson-fermion) scattering length. With our experimental parameters  $\mu_{bf}\sim \frac{1}{7}\mu_{bb}$. $\mu_{bf}$ is uniform across the BEC (since the size of the fermions cloud is larger due to Fermi pressure) and thus should not change the density.
During the release from the trap the boson-fermion interaction energy is divided between bosons and fermions as the clouds expand. To verify that it does not introduce a significant systematic error to our density evaluation in the presence of fermions, we compared the chemical potential of BEC’s with similar atom number (determined from absorption imaging according to Beer-Lambert law) with and without fermions (Figure \ref{fig2}). We did not observe a change in measured chemical potential after expansion in the presence of fermions.

\subsection*{Boson-Fermion frequency shift}
\begin{figure}[tb]
   \centering
    \includegraphics[width=0.75\textwidth]{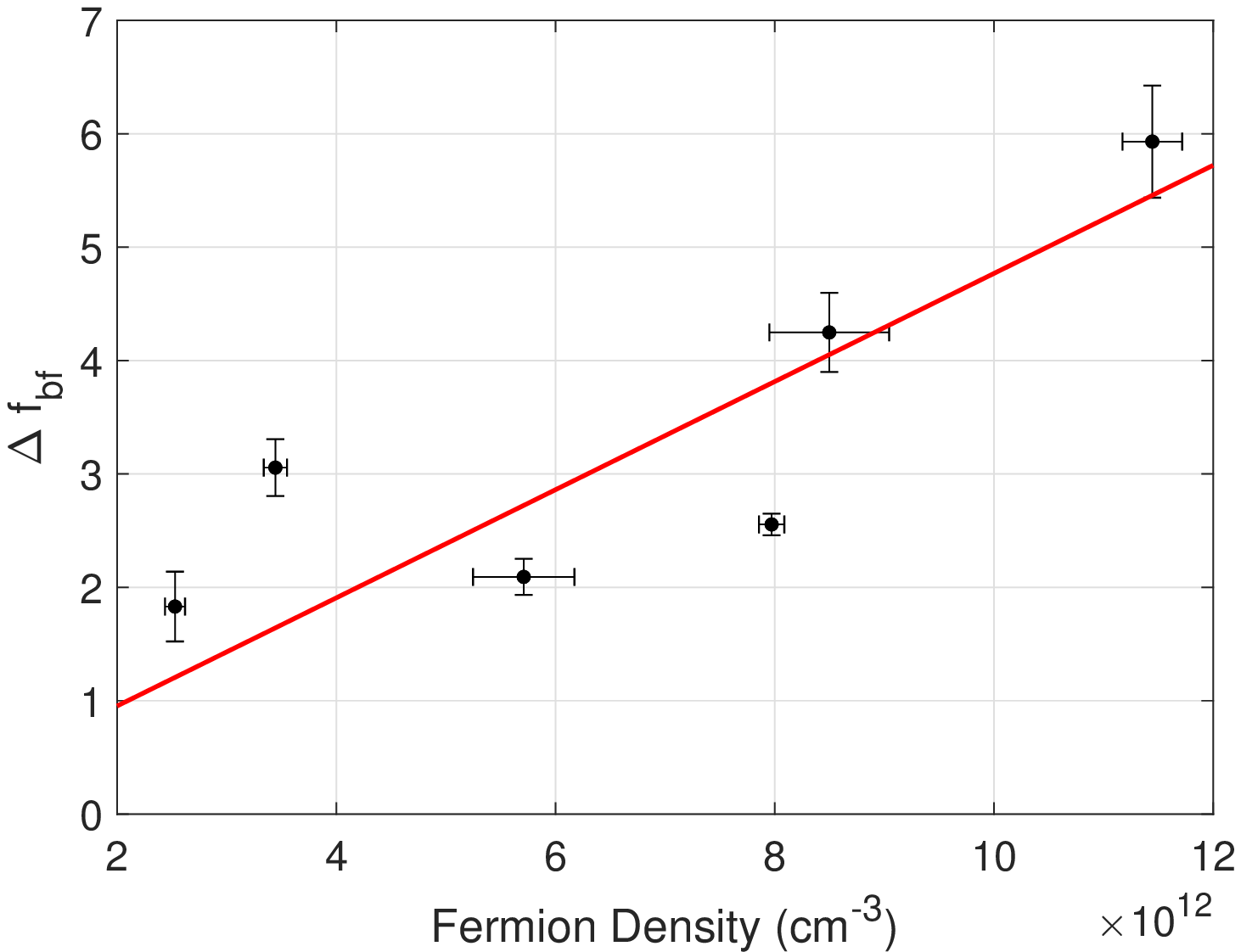}
    \caption{\label{fig3} Boson-Fermion frequency shift for different fermion densities. We measured the difference in frequencies with and without fermions for thermal mixture of bosons and fermions.}
\end{figure}

The fermion density in our measurements is $(8.6\pm 2) \times10^{12}$ cm$^{-3}$, making it hard to see the effect of fermion density on mediated interactions or on boson-fermion interaction. We have measured the boson-fermion frequency shift $\Delta f_{bf}$ with thermal clouds of bosons and fermions where it is easier to change the density of fermions and mediated interactions are negligible. We measured the difference in frequency of the bosons with and without fermions at different fermion densities. The change is only due to boson-fermion frequency shift, since the boson density is ~20 times smaller than in the BEC and the mediated interaction effect is weak and below our resolution. In this measurement (Figure \ref{fig3}) we get $a_{bf_2}-a_{bf_1}=-3.9 \pm 0.5^{stat} \pm 2^{sys} ~a_0$, in agreement with our results for degenerate Bose-Fermi mixture. The systematic uncertainty is from estimating the fermion density.

\subsection*{Systematic Errors}

Our frequency is generated from a commercially bought oscillator (Anritzu 68347B) locked to 10 MHz oscillator (SRS FS725) with $2\times10^{-12}$ stability after 100 s, which would be a 0.014 Hz error in our frequency, much below our measured frequency shifts. 

Our measurements are done in a constant  magnetic field of $\sim0.19$ G where the sensitivity to magnetic field on the clock transition is $\sim0.2$ Hz/mG. Using a magnetic sensitive MW transition, we measured our change in DC magnetic field over 10 hours to be $<1$ mG, and the 50 Hz noise felt by the atoms to be $1.1$ mG peak to peak.  Another source of frequency noise is light shift. 

In our trap we measured a light shift of $\sim 23$ Hz per Watt in both beams. We operate at $\sim0.5$ Watt per beam and stabilize the power in the beams to $<3\%$ which would cause a $0.3$ Hz shift. These effects should be averaged over time since we change the measurement parameters randomly (boson densities, Ramsey times, with/without fermions).

We repeat our measurements for different Ramsey times, different boson densities and with/without fermions. The effect of mediated interactions is given by $\eta=\left(\frac{\mathrm{d}\Delta{f}}{\mathrm{d}n_b}\right)_{b+f}/\left(\frac{\mathrm{d}\Delta{f}}{\mathrm{d}n_b}\right)_b = 1+\frac{\Delta f_{med}}{\Delta f_b}$, any systematic error that is common to measuring with/without fermions will not affect it. It is only weakly sensitive to errors in fermion density, $\eta-1 \propto n_f^{1/3}$.

\begin{figure}[tb]
   \centering
   \includegraphics[width=\textwidth]{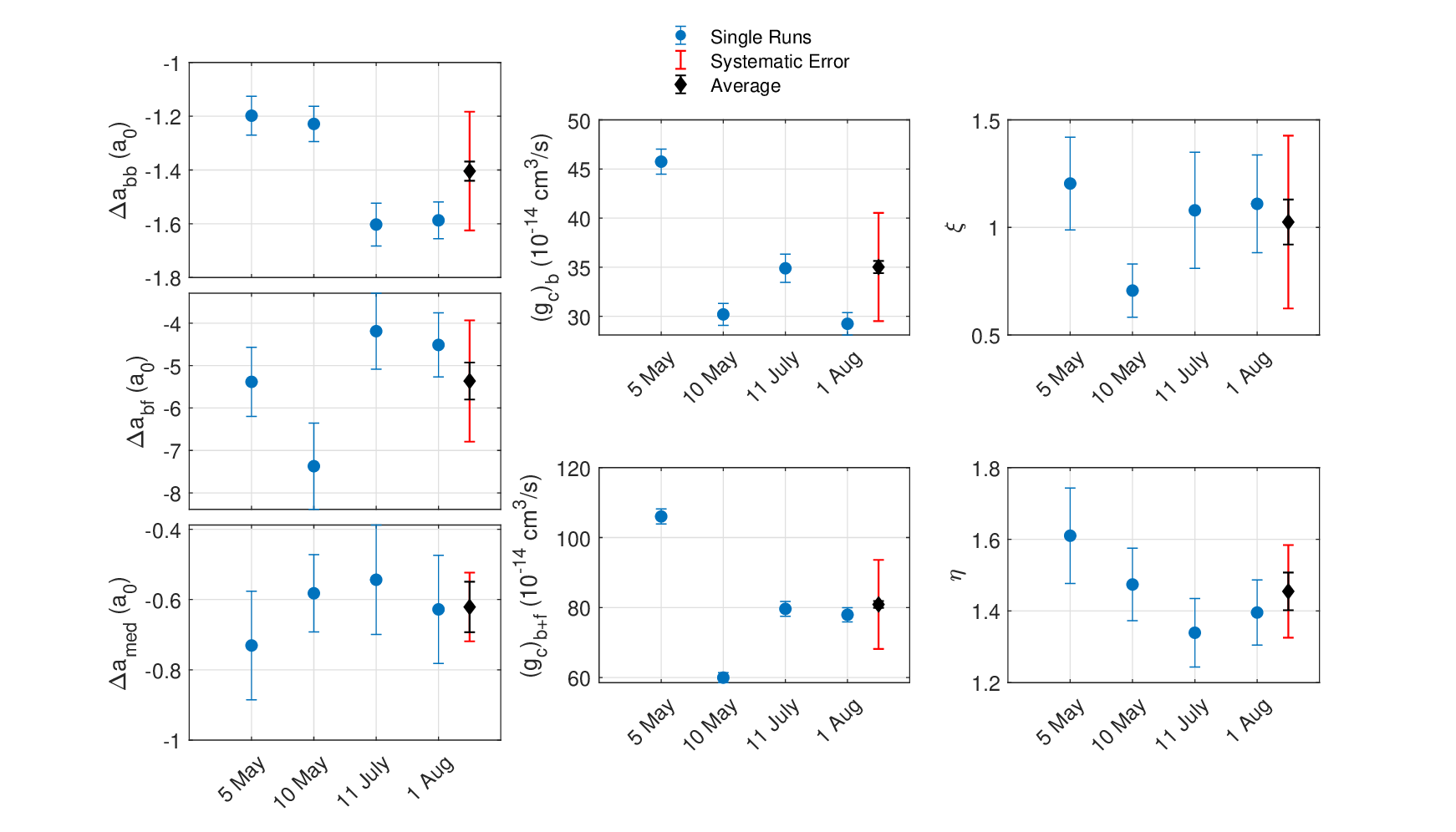}
    	\caption{Results from Four Experimental Runs for different parameters. Each run has $\sim 3000$ data points. Data is taken during $\sim 50$ hours in each run. Filled circles (blue) are the results from each run, with one standard deviation. Filled diamond (black) the mean of the four runs with statistical error in black and estimated systematic error in red. Start date for each run is presented in the horizontal axis, all in 2019.}
		\label{fig:combined_data}
\end{figure}

All the results reported in the paper were taken from four long runs of the experiment, each run has $\sim 3000$ data points and lasts $\sim 50$ hours, acquiring the data. The summary of the data from the different runs is presented in Figure \ref{fig:combined_data}. Filled circles (blue)  are the results of each run, with one standard deviation error taken from a fit to eq. (2). The result of a weighted mean of all runs is shown (diamond). The main source of systematic error is from measuring the densities of the clouds, which is hard to estimate. We used our measurements for boson-boson and boson-fermion frequency shifts ($\frac{\mathrm{d}\Delta{f_bb}}{\mathrm{d}n_b}$ and $\frac{\mathrm{d}\Delta{f_bf}}{\mathrm{d}n_f}$ respectively) to estimate a systematic error of $16\%$ for boson density and $27\%$ for fermion density. Assuming that the error from one data set to another is systematic, since each run has many measurements and there is a long time between the data sets.

